\documentclass{ifacconf}

\usepackage{enumitem}                           
\usepackage {xcolor}
\usepackage{float}
\usepackage{cuted}

\usepackage{graphicx}      
\usepackage{natbib}        
\usepackage{booktabs}
\usepackage{changes}
\usepackage{cancel}
\usepackage{hhline}           

\usepackage{tikz}
\usetikzlibrary{shapes,arrows}
\usepackage{amssymb,amsmath,graphicx,multicol,epstopdf}
\usepackage{comment}
\usepackage{multirow}

\newtheorem{Remark}{Remark}[section]
\newtheorem{Corollary}{Corollary}[section]

\newtheorem{Theorem}{Theorem}[section]
\newtheorem{Proposition}{Proposition}[section]

\newtheorem{Lemma}{Lemma}[section]
\newtheorem{Assumption}{Assumption}[section]
\newcommand{\floor}[1]{\left\lfloor #1 \right\rfloor}

\DeclareMathOperator*\argmin{arg\,min}
\DeclareMathOperator*\argmax{arg\,max}

\begin{document}
\begin{frontmatter}

\title{A Local Gaussian Process Regression Approach to Frequency Response Function Estimation\thanksref{footnoteinfo}} 

  \thanks[footnoteinfo]{ This work was funded by NSFC under contract No. 62273287, Shenzhen Science and Technology Innovation Commission under contract No. JCYJ20220530143418040, and the Thousand Youth Talents Plan funded by
the central government of China.} 

\author{Xiaozhu Fang}, 
\author{Yu Xu},
\author{Tianshi Chen}

\address{School of Data Science and Shenzhen Research Institute of Big Data, The Chinese University of Hong Kong, Shenzhen 518172, China (e-mail: \{xiaozhufang,yuxu19\}@link.cuhk.edu.cn, tschen@cuhk.edu.cn).}

\begin{abstract}                
Frequency response function (FRF) estimation is a classical subject in system identification. In the past two decades, there have been remarkable advances in developing local methods for this subject, e.g., the local polynomial method, local rational method, and iterative local rational method. The recent concentrations for local methods are two issues: the model order selection and the identification of lightly damped systems. To address these two issues,  we propose a new local method called local Gaussian process regression (LGPR). We show that the frequency response function locally is either analytic or resonant, and this prior knowledge can be embedded into a kernel-based regularized estimate through a dot-product kernel plus a resonance kernel induced by a second-order resonant system. The LGPR provides a new route to tackle the aforementioned issues. In the numerical simulations, the LGPR shows the best FRF estimation accuracy compared with the existing local methods, and moreover, the LGPR is more robust with respect to sample size and noise level.  
\end{abstract}


\end{frontmatter}

\section{Introduction}
Frequency response function (FRF) estimation in the frequency domain is a classic subject in the field of system identification, see e.g. \cite{LJUNG:99, PS12}. The early methods for FRF estimation include the classic spectral estimation method and the empirical transfer function estimate (ETFE), see e.g. \cite{LJUNG:99}. In the past two decades, there have been remarkable advances in developing new methods for this subject, see \cite{SGS18} for a review. Among these methods, the local polynomial method (LPM) \citep{SVBP09}, local rational method (LRM) \citep{MG12}, and iterative local rational method (ILRM) \citep{Geerardyn16} have gained significant attention due to their capability to almost reject the leakage. The pros and cons of the above local methods have been debated recently \citep{SGS18, PPVL20}.

For the LPM, since polynomials are employed as basis functions within each sliding window, the model order selection issue is unavoidable. The order of polynomials is often set to be 2 \citep{SGS18}, or tuned by minimizing a criterion, e.g., the modified minimum
description length (MDL) \citep{PPVL20}. Moreover, the identification of lightly damped systems is a critical issue for the LPM because the FRFs of lightly damped systems have resonance peaks, which are not smooth and are hard to capture by the LPM if the number of frequency bins is not much within the sliding window. In contrast, the LRM often gives better performance than the LPM by using a local rational model within each sliding window. The LRM also has the model order selection issue, and additionally, its noise variance estimate is biased due to the Levy approximation, especially for a low signal-to-noise ratio (SNR) \citep[Guideline 3.1]{Geerardyn16}.  As for the ILRM, it has less bias in the noise variance estimate than the LRM, but it involves nonconvex optimizations with severe local minima issues \citep{VS19}.  The success of the ILRM highly depends on the dedicated model order selection and the starting point of the optimization, where \cite{PPVL20} suggested using the result of LRM as the starting point.  To summarize,  none of the existing local methods has shown consistent superiority over others. 

In contrast with the above local methods, the Gaussian process regression (GPR) is a global method that makes use of all data points in the FRF estimation. Moreover, the GPR is a kernel-based nonparametric method: on the one hand, the prior knowledge about the FRF  to be estimated can be embedded in the estimated FRF by designing a suitable kernel, e.g., \cite{LC16}, and on the other hand, the model complexity of the estimated FRF can be tuned continuously by tuning the hyper-parameters through e.g. the empirical Bayes method. 

In this paper, we are interested in developing a new local method inspired by the global GPR. 
More specifically,

\begin{itemize}
\item we first propose a kernel-based regularized version of the LPM and call it the local regularized polynomial method
(LRPM). If the system is not lightly damped, then 
the coefficients of the polynomial basis functions are absolutely summable, and then the  diagonal (DI) kernel in \cite{COL12} is introduced to embed this prior knowledge;
\item we then show that the LRPM can be extended to a more general local GPR (LGPR). Particularly, we show that,  for lightly damped systems, since their FRFs locally can be either analytic or resonant, a dot-product (DP) kernel \citep{Smola00} and a resonance kernel induced by a second-order resonant system can be designed to incorporate such prior knowledge, respectively. 

\item we finally demonstrate the validity of the LGPR with the DP plus one resonance (DPpR1) kernel through numerical simulations, compared with the existing local methods. 
\end{itemize}

The remaining part of this paper is organized as follows. In Section \ref{se:problem}, FRF estimation and the LPM are briefly introduced. The main derivations of the LRPM and LGPR are shown in Section \ref{se:lrpm} and  Section \ref{se:lgpr}, respectively. In Section \ref{se:sim}, the proposed method is tested by simulations.

\section{Background and Problem Formulation}\label{se:problem}

\subsection{Frequency Response Function Estimation}
We consider the causal, stable, continuous-time, single-input single-output, and linear time-invariant system with output disturbed by filtered white noise: 
{\small\begin{equation}
    \begin{aligned}
y(t)&= \int_{0}^\infty u(t-\tau)g(\tau)d\tau+v(t),\\
v(t)&= \int_{0}^\infty e(t-\tau)h(\tau)d\tau, 
\end{aligned}\label{eq:sys}
\end{equation}}\normalsize
where $t\geq 0$ is the time index, $u(t),y(t), v(t)\in \mathbb{R}$ are the input, output, and disturbance, respectively,  $g(\tau), h(\tau)\in \mathbb{R}$ is the impulse response of the system and noise model, respectively, and $e(t)\in \mathbb{R}$ is assumed to be the unobserved white Gaussian noise. 

Let the input and output be sampled at $\{u(t), y(t)\}$, $t=0,T_s,2T_s,\cdots, (N-1)T_s$ with $N$ being the sample size and $T_s$ being the sampling interval. The input $u(t)$ is assumed band-limited under the Nyquist frequency $\pi/T_s$, or, otherwise, the aliasing error should be modeled together with the transient error later.  Denote
 $U(k)$, $Y(k)$, $V(k)$ and $E(k)$  the $N$-point discrete Fourier transform ($N$-point DFT) of $u(t)$, $y(t)$, $v(t)$, and $e(t)$, respectively, at frequency bin $k$. That is, if we let $x(t)$ represent $u(t),y(t), v(t)$ or $e(t)$, and $X(k)$ its corresponding $N$-point DFT, then it holds that {\small\begin{align}
&X(k)=\frac{1}{\sqrt{N}}\sum\limits_{n=0}^{N-1}x(nT_s)e^{-\frac{2\pi jn k}{N}}. \label{eq:np_dft}
\end{align} }\normalsize
For white Gaussian noise $e(t)$, $E(k)$ is independent over $k$, circular complex Gaussian distributed, see e.g.,\citep[Sec. 7.2.1]{PS12}. 

The frequency domain correspondent of \eqref{eq:sys} is 
{\small\begin{equation}\label{eq:sys_freq}
    \begin{aligned}
&Y(k)= G(j\omega_k)U(k)+ T(j\omega_k)+V(k),\\
&V(k)= H(j\omega_k)E(k), \ \omega_k =2\pi k/NT_s,\ k=0,1,\cdots, \floor{N/2},
\end{aligned} 
\end{equation}}\normalsize
where $\floor{N/2}$ denotes the maximum integer no larger than $N/2$, $G(j\omega)$ and $H(j\omega)$ are the system and noise transfer functions, respectively, and $T(j\omega_k)$ denotes the transient term (aka. leakage) incorporating the system transient and noise transient. The derivation of \eqref{eq:sys_freq} is skipped and refers to \cite{PS12}. 

The goal is to obtain the estimates of the frequency response function (FRF) $G(j\omega_k)$  and  noise variance  $\sigma^2_k: = \text{Cov}(V(k))$  from the input-output spectra $U(k), Y(k)$ in the interested frequency band.
\subsection{Existing Local Methods} 
Recently, it started to be clear that there is plenty of structure in  $T(j\omega_k)$, and some local methods have been proposed for the FRF estimation, see \cite{SGS18}. Specifically, the local methods use low-order parametric models $G_k(j\omega;\theta_k)$ and $T_k(j\omega;\theta_k)$ to approximate  $G(j\omega)$ and  $T(j\omega)$ in the local window, 
\begin{align*}
[j\omega_{k-\ell},\cdots,  j\omega_{k-1}, j\omega_{k}, j\omega_{k+1}, \cdots,j\omega_{k+\ell}],    
\end{align*}
where $j\omega_k$ is any excited frequency in \eqref{eq:sys_freq}, $\theta_k\in \mathbb{R}^{n_k}$ being the parameters to be estimated, and $n_k\in \mathbb{N}$ being the dimension of $\theta_k$. 
Then the predicted output in the local window is given by, with $r = -\ell, \ell+1, \cdots, \ell$, 
{\small\begin{align}\label{eq:local_pred}
  \hat{Y}(k+r;\theta_k)= G_{k}(j\omega_{r}; \theta_k)U(k+r)+T_{k}(j\omega_{r}; \theta_k).
  \end{align}}\normalsize
Then the parameters $\theta_k$ can be estimated from the local data $Z_k$ by minimizing an estimation criterion $\mathcal{V}$,  
 {\small \begin{subequations}\label{eq:vector_u_y}
        \begin{align}
&\hat{\theta}_k = \underset{\theta_k}{\argmin}\  \mathcal{V}( \theta_k, Z_k ),\quad  Z_k=\{U(k+r), Y(k+r)\}_{r=-\ell}^{\ell}. \nonumber
  \end{align}
  \end{subequations}}\normalsize
Finally,  the central of the estimated local models is remained for the final estimate, i.e., 
{\small\begin{align}\label{eq:final_local_est}
&\hat{G}(j\omega_k)=G_{k}(0;\hat \theta_k ),\quad \hat{T}(j\omega_k)=T_{k}(0;\hat \theta_k ),
\end{align}}\normalsize
and  the noise variance estimates are given by
{\small\begin{align}
    \hat{\sigma}^2_k = \frac{1}{2\ell+1-n_k} \sum\limits_{r=-\ell}^\ell |Y(k+r)- \hat Y(k+r;\hat{\theta}_k)|^2. \label{eq:noise_est}
\end{align}}\normalsize
The existing local methods use different local models and estimation criteria. 
The local polynomial method (LPM) chooses polynomials as the local models and squared local output error (LOE) as the estimation criterion: 
{\small
\begin{align}
&\left\{\begin{array}{lll}
 G_k( j\omega_{r};\theta_k )=B^c_k(j\omega_r) ,\quad T_k( j\omega_{r};\theta_k )= I^c_k(j\omega_r),\\
    \theta_k=[b^c_{0,k}, \cdots, b^c_{n_{b,k},k}, i^c_{0,k}, \cdots, i^c_{n_{i,k},k} ]^T , \\
    n_k= n_{b,k}+ n_{i,k}+ 2,
\end{array}\right. \label{eq:polynomial_model}\\
&\mathcal{V}_{\text{LOE}}(\theta_k, Z_k)= \sum\limits_{r= -\ell}^\ell |Y(k+r)- \hat Y(k+r;\theta_k)|^2,  \label{eq:loe_output}
\end{align}}\normalsize
where  $X^c_k(j\omega)$ with $X=B, A, I$ denote the polynomials with parameters $x_{n,k}^c$ and polynomial order $n_{x,k}$
{\small\begin{align}\label{eq:polynomials}
    X^c_k(j\omega)= \sum\limits_{n=0}^{n_{x,k}} x^c_{n,k} (\alpha \omega)^n, \text{ with } x_{n,k}^c, \alpha \in \mathbb{C}, 
\end{align}}\normalsize
with $\alpha$ being a constant. 
The local rational method (LRM)  chooses rational functions as the local models and the local levy (LL) method as the estimation criterion:   
{\small
\begin{align}
&\left\{\begin{array}{lll}
     G_k( j\omega_{r};\theta_k )= \frac{B^c_k(j\omega_r)}{A^c_k(j\omega_r)} ,\quad T_k( j\omega_{r};\theta_k )= \frac{I^c_k(j\omega_r)}{A^c_k(j\omega_r)}, \\
    \theta_k\!=\![a^c_{1,k}, \!\cdots\!, a^c_{n_{a,k},k}, b^c_{0,k},\! \cdots\!, b^c_{n_{b,k},k}, i^c_{0,k}, \!\cdots\!, i^c_{n_{i,k},k} ]^T, \\
    n_k\!=\! n_{a,k}+ n_{b,k}+ n_{i,k}+ 2, \quad a_{0,k}^c\!=\!1,
    \end{array}\right. \label{eq:rational_model}\\
        &\mathcal{V}_{\text{LL}}( \theta_k, Z_k)\!=\! \sum\limits_{r= -\ell}^\ell |A^c_k(j\omega_r)|^2  |Y(k+r)\!-\! \hat Y(k+r;\theta_k)|^2.\label{eq:ll_output}
\end{align}}\normalsize
The iterative local rational method (ILRM) chooses rational functions \eqref{eq:rational_model}  as the local models and the LOE method \eqref{eq:loe_output} as the estimation criterion.

The comparison of the above local methods has been discussed in \cite{VS19, SGS18, PPVL20}. 
In short, the LPM has difficulties in identifying lightly damped systems, while the LRM and ILRM do not. Moreover,  the LRM has a biased noise variance estimate due to the LL method \eqref{eq:ll_output}, while the ILRM involves a nonconvex optimization and suffers from severe local minima issues.  It was suggested in \cite{PPVL20}  to use the result of the LRM as the starting point of the optimization for the ILRM. 
\subsection{Problem Statement} 
The model order selection is essential for the existing local methods. How to choose the polynomial orders can be heuristic, e.g. $n_{a,k}, n_{b,k}$, and $n_{i,k}$ are suggested to be $2$ \citep{SGS18}, or the polynomial orders can be tuned by minimizing a criterion, e.g. modified minimum description length (MDL) \citep{PPVL20} as follows: 
\begin{align}\label{eq:mdl}
\underset{n_{a,k}, n_{b,k}, n_{i,k} }{\argmin} \quad \hat{\sigma}^2_k  e^{\log(4\ell+2)\frac{n_k}{2\ell+1-n_k -2}}
\end{align} 
where $n_{a,k} =0$ for the LPM.  
In this paper, we first develop a kernel-based regularized version of the LPM and call it the local regularized polynomial method (LRPM). The LRPM can tune the model order in a continuous manner.  

On the other hand, the LRPM also has difficulties in identifying lightly damped systems; to tackle that, we propose the local Gaussian process regression (LGPR), as an extension of the LRPM. 
\section{Local Regularized  Polynomial Method} \label{se:lrpm}
Since the LPM is a linear model with respect to $\theta_k$, it can be reformulated in matrix-vector form:
{\small\begin{align}\label{eq:linear_regre_model}
&Y_k= \Psi_k\theta_k+ V_k,
\end{align}}\normalsize
with  
{\small\begin{align*}
&U_k= [ U(k-\ell), \cdots,  U(k+\ell)]^T, \quad Y_k= [ Y(k-\ell), \cdots, Y(k+\ell)]^T,\\
&V_k= [ V(k-\ell), \cdots, V(k+\ell)]^T, \quad \phi^n = [(\alpha\omega_{-\ell})^n, \cdots , (\alpha\omega_{\ell})^n]^T,  \\
&\Phi^{n}= [\phi^0, \cdots, \phi^n ]\in \mathbb{R}^{(2\ell+1)\times (n+1)},\ n = 0,1,\cdots\\
&\Psi_k =[ \text{diag}[U_k] \Phi^{n_{b,k}} \ \Phi^{n_{i,k}} ]\in \mathbb{C}^{(2\ell+1)\times n_k }
\end{align*}}\normalsize
where $T$ denotes the transpose, and diag$[U_k]$ denotes the diagonal matrix with its diagonals being $U_k$.
\subsection{Bayesian/Regularized Estimate}
Assume that $\theta_k$ in \eqref{eq:linear_regre_model} is a complex Gaussian random vector (CGRV) with mean zero, covariance matrix $\Gamma_{\theta_k}$, and relation matrix $C_{\theta_k}$,  and $V_k$ is an i.i.d. and  circular CGRV with variance $\sigma_k^2$, independent with $\theta_k$, denoted by 
{\small\begin{align}
&\theta_k\sim  \mathcal{CN}(0,\Gamma_{\theta_k}, C_{\theta_k}),\quad V_k\sim  \mathcal{CN}(0,\sigma_k^2 I_{2\ell+1}, 0 ), \label{eq:prior_V}
\end{align}}\normalsize
where $I_{n}$ is a $n\times n$ identity matrix. 
Then we have the maximum a posteriori (MAP) estimate as follows. 
\begin{Proposition}[MAP Estimate for $\theta_k$]\label{pr:map_est}
Consider \eqref{eq:linear_regre_model}\\
with  \eqref{eq:prior_V}, where $V_k$ is independent of $\theta_k$. Given the observation  $\widetilde{Y}_k =[Y_k^T\ Y_k^H]^T$,  the MAP estimate of  $\widetilde{\theta}_k = [\theta_k^T\ \theta_k^H]^T$ is 
{\small\begin{align}\label{eq:map_theta}
&\widehat{\widetilde{\theta}}_k^{\text{MAP}}= M_{\theta_k} \widetilde{\Psi}_k (\widetilde{\Psi}_k M_{\theta_k} \widetilde{\Psi}_k^H+ \sigma_k^2 I_{4\ell+2})^{-1}\widetilde{Y}_k,  \\
& M_{\theta_k}= \left[
\begin{matrix}
\Gamma_{\theta_k} & C_{\theta_k}\\
\overline{C_{\theta_k}}& \overline{\Gamma_{\theta_k}} \\
\end{matrix}
\right], 
  \widetilde{\Psi}_k= \left[
\begin{matrix}
\Psi_k & 0\\
0 & \overline{\Psi}_k\\
\end{matrix}
\right]. \nonumber
\end{align}}\normalsize
where $H$ denotes the Hermitian transpose. 
\end{Proposition}
Likewise to \cite[Prop. 13]{PDCDL14},  \eqref{eq:map_theta} can be associated to a regularized least square (RLS) estimate 
{\small\begin{align}
\widehat{\widetilde{\theta}}_k^{\text{MAP}}&=  \underset{\widetilde{\theta_k}}{\argmin} \|\widetilde{Y}_k-\widetilde{\Psi}_k\widetilde{\theta}_k   \|^2 + \sigma_k^2 \widetilde{\theta}_k ^HM_{\theta_k} ^{-1} \widetilde{\theta}_k, \nonumber\\
\hat{\theta}_k^{\text{MAP}} & = \underset{\theta_k}{\argmin}\  V_{\text{LOE}}(U_k, Y_k; \theta_k )+ \frac{1}{2}\sigma_k^2 \widetilde{\theta}_k ^HM_{\theta_k} ^{-1} \widetilde{\theta}_k. \label{eq:rls_est_theta}
\end{align}}\normalsize
Now we obtain a new local method, namely the LRPM,  which replaces the LS estimate with the RLS estimate \eqref{eq:rls_est_theta} in the LPM. 
 The key issue for the LRPM is the kernel design, i.e., determining the parameterization of kernel matrix $M_{\theta_k}$ by hyperparameters $\eta_k\in \mathbb{R}^p$.
\subsection{Kernel Design for the LRPM}\label{se:kernel_design}
Let $M_{\theta_k} (\eta_k)$ denote the parameterized kernel matrix to be designed, and it must be positive semidefinite in the complex-valued sense. However, this positive semidefinite condition is hard to fulfill in the design of $\Gamma_{\theta_k}(\eta_k)$ and $C_{\theta_k}(\eta_k)$, and thus we consider the equivalent way to describe the CGRV $\theta_k$, 
\small \begin{align*}
 & \left[
\begin{matrix}
\mathfrak{R}[\theta_k]\\
\mathfrak{I}[\theta_k]
\end{matrix}
\right]\sim \mathcal{N}(0,K_{\theta_k}(\eta_k)), \quad K_{\theta_k}(\eta_k)= \left[
\begin{matrix}
K_{\theta_k,rr}& K_{\theta_k,ri}\\
K_{\theta_k,ir}& K_{\theta_k,ii}
\end{matrix}
\right],
 \end{align*}\normalsize
 where $\mathcal{N}$ denotes the real-valued Gaussian distribution,  $\mathfrak{R}[\theta_k]$ and  $\mathfrak{I}[\theta_k]$ denote the real and imaginary components of $\theta_k$, respectively, $K_{\theta_k}(\eta_k)$ is the composited covariance matrix consisting with e.g. $K_{\theta_k,rr}= \mathbb{E}[\mathfrak{R}[\theta_k]\mathfrak{R}[\theta_k]^T]$, and  $M_{\theta_k}(\eta_k)$ and $K_{\theta_k}(\eta_k)$ follow the following relationship:
\small\begin{align*}
M_{\theta_k} (\eta_k)&=\left[
\begin{matrix}
I_{n_k}& j I_{n_k}\\
I_{n_k}& -j I_{n_k}
\end{matrix}
\right] K_{\theta_k} (\eta_k)\left[
\begin{matrix}
I_{n_k}& j I_{n_k}\\
I_{n_k}& -j I_{n_k}
\end{matrix}
\right]^H.
\end{align*} \normalsize
Therefore, the task of kernel design now becomes the parametrization of the real-valued kernel matrix $K_{\theta_k} (\eta_k)$.
\subsubsection{Finding True $\theta_k$}
The task of kernel design now becomes the parametrization of the real-valued kernel matrix $K_{\theta_k} (\eta_k)$.
However, the true $\theta_k$, denoted by $\theta_k^o=[b^o_{0,k}, \cdots, b^o_{n_{b,k},k}$, $i^o_{0,k}, \cdots, i^o_{n_{i,k},k} ]^T$, is unknown. From the experience of stable kernel design \cite{PDCDL14}, it is necessary to ensure that $\theta_k^o$ is convergent as $n_{b,k}, n_{i,k}\rightarrow \infty$, and both $b^o_{n_{b,k},k}$ and $ i^o_{n_{i,k},k}$ are decaying to zero. Unfortunately, such convergence may not exist for arbitrary true system and transient, denoted by $G^o(j\omega)$ and $T^o(j\omega)$, respectively, and we thus need the following assumption.

\begin{Assumption}\label{as:analytic}
Assume that, for given $\omega_\ell$ and $\omega_k$, $G^o(j\omega)$ is analytic in the disk $\{j\omega\in \mathbb{C}: |\omega-\omega_k|<\omega_\ell \}$.  
\end{Assumption}
Then we have the power series expansion for the local FRF. 
\begin{Theorem}\label{th:true_theta}
If $G^o(j\omega)$ is analytic in the disk $\{j\omega\in \mathbb{C}: |\omega-\omega_k|<\omega_\ell \}$,  then it has a unique  power series expansion,
{\small\begin{align*}
G^o(j\omega_k+ j\omega)&= \sum\limits_{n=0}^{\infty} \bigg(  \frac{(G^o)^{(n)}(j\omega_k)}{n!} \bigg) (j\omega)^n, 
\end{align*}}\normalsize
converges uniformly for $|\omega|< \omega_\ell$. Associated with polynomials \eqref{eq:polynomials} in the LPM, it holds that, for any $0<\omega_r<\omega_\ell$, 
{\small\begin{align}
b^o_{n,k}& \triangleq   \frac{(G^o)^{(n)}(j\omega_k)}{n!}(\frac{j}{\alpha})^n ,\label{eq:opttheta}
\\
|b^o_{n,k}|& \leq (\frac{1}{\alpha \omega_r})^n \sup \limits_{{|\omega-\omega_k|}=\omega_r}| G^o(j\omega)|. \label{eq:stability_theta}
\end{align}}\normalsize
\end{Theorem}
\noindent \textit{Proof}:  An immediate result of  \cite{GAMELIN03}[p. 144].

The last inequality \eqref{eq:stability_theta}  results in the following corollary. 
\begin{Corollary}\label{co:stable_theta}
If $G^o(j\omega)$ is analytic in the disk $\{j\omega\in \mathbb{C}: |\omega-\omega_k|<\omega_\ell \}$ and $|\alpha\cdot \omega_\ell|>1$, then the sequence $\{b^o_{n,k}\}_{n=0}^\infty$ is stable, i.e., $\sum_{n=0}^\infty |b^o_{n,k}|<\infty$.
\end{Corollary}
This stability enables the kernel design as follows, and it can also apply to $\{i^o_{n,k}\}_{n=0}^\infty$ in the same manner. 

\subsubsection{Diagonal Kernel }
Given the stability of $\theta^o_k$,  we propose the diagonal (DI) kernel \citep{COL12} for the LRPM:
{\small\begin{align}
&K_{\theta_k}^{\text{DI}}(\eta_k):\left\{
\begin{array}{lll}
&K_{\theta_k, rr}= \text{diag}([\alpha_G , \alpha_G \lambda_G^1,  \cdots, \alpha_G \lambda_G^{n_{b,k}}, \\
&\quad \alpha_T , \alpha_T \lambda_T^1, \cdots, \alpha_T \lambda_T^{n_{i,k}}])\\
&K_{\theta_k, ii}= \text{diag}([\beta_G , \beta_G \kappa_G^1, \cdots, \beta_G \kappa_G^{n_{b,k}}, \\
&\quad \beta_T ,\beta_T \kappa_T^1,  \cdots, \beta_T \kappa_T^{n_{i,k}}])\\
&K_{\theta_k, ir}= 0;
\end{array}\right.\nonumber\\
&\alpha_G, \alpha_T,\beta_G, \beta_T\geq 0, \lambda_G, \lambda_T,\kappa_G, \kappa_T\in [0,1/\alpha \omega_r ), \label{eq:di_kernel} \\
&\eta_k=[\alpha_G,  \lambda_G, \alpha_T, \lambda_T, \beta_G,  \kappa_G, \beta_T, \kappa_T]. \nonumber
\end{align}}\normalsize
Moreover, there exists the structural resemblance between $G^o( j\omega)$ and $ T^o( j\omega)$, see e.g. \cite[Sec. 6.3.2]{PS12} and \cite[Sec 5.3]{LC16}. Then it is assumed that 
{\small\begin{align}\label{eq:trans_assup}
\lambda_T= \lambda_G, \quad \kappa_T= \kappa_G,\quad  \alpha_G/\alpha_T =\beta_G/\beta_T .
\end{align}}\normalsize
From \eqref{eq:di_kernel} and \eqref{eq:trans_assup},  we obtain the corresponding  $M_{\theta_k}(\eta_k)$, 
{\small\begin{align}
&M_{\theta_k}^{\text{DI}}(\eta_k)\!:\!\left\{
\begin{array}{lll}
&\Gamma_{\theta_k}\!=\! \text{diag}([\alpha_G\!+\!\beta_G , \alpha_G \lambda^1\!+\! \beta_G \kappa^1, \cdots, \alpha_G \lambda^{n_{b,k}}\!+\!\\
&\beta_G \kappa^{n_{b,k}}, \alpha_T \!+\! \beta_T ,\alpha_T \lambda^1\!+\! \beta_T \kappa^1,  \cdots, \alpha_T \lambda^{n_{i,k}}\!+\!\beta_T \kappa^{n_{i,k}}])\\
&C_{\theta_k}\!=\! \text{diag}([\alpha_G\!-\!\beta_G ,\alpha_G \lambda^1\!-\! \beta_G \kappa^1, \cdots, \alpha_G \lambda^{n_{b,k}} \!-\!\\
&\beta_G \kappa^{n_{b,k}}, \alpha_T\!-\! \beta_T ,\alpha_T \lambda^1\!-\! \beta_T \kappa^1, \cdots, \alpha_T \lambda^{n_{i,k}}\!-\!\beta_T \kappa^{n_{i,k}}]) \\
\end{array}\right.\nonumber\\
& \eta_k=[\alpha_G, \alpha_T, \lambda, \beta_G, \kappa],\quad  \beta_T=  \alpha_T*\beta_G/\alpha_G. \label{eq:M_di}
\end{align}
}\normalsize
\subsection{What if Assumption \ref{as:analytic} is invalid?}\label{se:light-damped}
For lightly damped systems, the resonance poles are too close to the imaginary axis, and thus Assumption \ref{as:analytic} is invalid at some $\omega_k$, and the power series in Theorem \ref{th:true_theta} do not converge.  Thus,  
the LRPM (and also the LPM) cannot be applied to lightly damped systems.

To tackle lightly damped systems, the polynomial models in the LPM are often replaced by rational function models, e.g., the LRM and ILRM.  As an alternative,  we 
extend the LRPM to the local Gaussian process regression (LGPR).

\section{Local Gaussian Process Regression}\label{se:lgpr}

The LGPR is straightforward by employing the GPR \cite[Thm. 3]{LC16} in the local window. Nonetheless, we prefer to derive the LGPR as an extension of the LRPM because this extension relationship would be important for the kernel design in Sec. \ref{subse:kernel_lrpm}. 

\subsection{The LGPR: an Extension of the LRPM }\label{subse:lgpr_from_lrpm}

It follows from Prop. \ref{pr:map_est} that $G_k=[G(j\omega_{-\ell}), \cdots, G(j\omega_\ell)]$ and $T_k=[T(j\omega_{-\ell}), \cdots, T(j\omega_\ell)]$ are also  CGRVs. Assume the independence of $G_k$ and $T_k$. Then the covariance matrix and relation matrix of $G_k$, denoted by $\Gamma_{G_k}$ and $C_{G_k}$, respectively, are given by 
{\small\begin{subequations}\label{eq:gamma_equivalent}
 \begin{align}
\Gamma_{G_k}\!=\!\Phi_k^{n_{b,k}} \Gamma^G_{\theta_{k}} (\Phi^{n_{b,k}}_k)^H, C_{G_k}\!=\!\Phi_k^{n_{b,k}} C^G_{\theta_{k}} (\Phi_k^{n_{b,k}})^T, \\
\Gamma_{T_k}\!=\!\Phi_k^{n_{i,k}} \Gamma^T_{\theta_{k}} (\Phi^{n_{i,k}}_k)^H, C_{T_k}\!=\!\Phi_k^{n_{i,k}} C^T_{\theta_{k}} (\Phi_k^{n_{i,k}})^T, 
\end{align}  
\end{subequations}}\normalsize
where $\Gamma^G_{\theta_{k}},  C^G_{\theta_{k}}\in \mathbb{C}^{(n_{b,k}+1)\times (n_{b,k}+1)}$ and $\Gamma^T_{\theta_{k}},  C^T_{\theta_{k}}\in \mathbb{C}^{(n_{i,k}+1)\times (n_{i,k}+1)}$ come from $\Gamma_{\theta_k}$ and $C_{\theta_k}$ with
{\small\begin{align*}
    &\Gamma_{\theta_k}= \left[\begin{matrix}
         \Gamma^G_{\theta_{k}}& 0\\
         0&  \Gamma^T_{\theta_{k}}
    \end{matrix}
    \right],  \ C_{\theta_k}= \left[\begin{matrix}
         C^G_{\theta_{k}}& 0\\
         0&  C^T_{\theta_{k}}
    \end{matrix}
    \right].
\end{align*}}\normalsize
Then we have the following proposition. 
\begin{Proposition}[MAP Estimate for $G_k$]\label{pr:map_est_G}
Consider \eqref{eq:sys_freq}\\ with $G_k \sim \mathcal{CN}(0, \Gamma_{G_k}, C_{T_k})$, $T_k \sim \mathcal{CN}(0,\Gamma_{G_k},C_{T_k})$, and $V_k$ in \eqref{eq:prior_V},  where $G_k, T_k, V_k$ are assumed independent of each other.  
Given the observation$\widetilde{Y}_k \triangleq [Y_k^T\ Y_k^H]^T$,  the MAP estimate of  $\widetilde{G}_k= [G_k^T\ G_k^H]^T$ is given by 
{\small\begin{align}
& \widehat{\widetilde{G}}_k^{\text{MAP}}\!=\! M_{G_k} \widetilde{\mathfrak{U}}_k (\widetilde{\mathfrak{U}}_k M_{G_k} \widetilde{\mathfrak{U}}_k^H+  M_{T_k}+\sigma_k^2 I_{4\ell+2})^{-1}\widetilde{Y}_k, \label{eq:map_est_G} \\
& \widehat{\widetilde{T}}_k^{\text{MAP}}\!=\! M_{T_k}  (\widetilde{\mathfrak{U}}_k M_{G_k} \widetilde{\mathfrak{U}}_k^H+  M_{T_k}+\sigma_k^2 I_{4\ell+2})^{-1}\widetilde{Y}_k, \label{eq:map_est_T} \\
& M_{G_k}\!=\! \left[
\begin{matrix}
\Gamma_{G_k} & C_{G_k}\\
\overline{C_{G_k}}& \overline{\Gamma_{G_k}} \\
\end{matrix}
\right], M_{T_k}\!=\! \left[
\begin{matrix}
\Gamma_{G_k} & C_{G_k}\\
\overline{C_{G_k}}& \overline{\Gamma_{G_k}} \\
\end{matrix}
\right],  \widetilde{\mathfrak{U}}_k\!=\!\text{diag}[U_k^T\ U_k^H], \nonumber
\end{align}}\normalsize
\end{Proposition}
Likewise,  $M_{G_k}$ and $M_{T_k}$ can be parametrized by hyperparamters $\eta_k$, denoted by $M_{G_k}(\eta_k)/M_{T_k}(\eta_k)$, which are estimated by the empirical Bayes (EB) method:  
{\small\begin{align}
&\hat{\eta}_k^{\text{EB}}, \hat{\sigma}_k^{\text{EB}}\!=\! \underset{\eta_k, \sigma_k}{\argmax}\  \frac{1}{2}\widetilde{Y}_k^H  O_k^{-1}(\eta_k)\widetilde{Y}_k + \frac{1}{2}\log |O_k(\eta_k)|,\nonumber\\
&O_k(\eta_k)= \widetilde{\mathfrak{U}}_k M_{G_k}(\eta_k)\widetilde{\mathfrak{U}}_k^H+ M_{T_k}(\eta_k)+ \sigma_k^2I_{4\ell+2}.\label{eq:eb}
\end{align}}\normalsize

\subsection{Kernel Design from the LRPM to the LGPR} \label{subse:kernel_lrpm}

The extension relationship between the LRPM and LGPR implies that all kernels for the LRPM can be transformed into the kernels for the LGPR via \eqref{eq:gamma_equivalent}.

\begin{Proposition}[Dot-Product Kernel]
Substituting the \\
DI kernel \eqref{eq:M_di} into \eqref{eq:gamma_equivalent}, we obtain
{\small\begin{align}\label{eq:ps_kerenl}
&\Gamma_{G_k}^{\text{DP}}\!=\! \sum\limits_{n=0}^{n_{b,k}} (\alpha_G\lambda^n\!+\! \beta_G\kappa^n) \phi^n(\phi^n)^H, \\
   &C_{G_k}^{\text{DP}}\!=\! \sum\limits_{n=0}^{n_{b,k}} (\alpha_G\lambda^n\!-\! \beta_G\kappa^n) \phi^n(\phi^n)^T 
\end{align}}\normalsize
where  $\phi^n$ is given in  \eqref{eq:linear_regre_model}, and  \eqref{eq:ps_kerenl} takes the form of the dot-product (DP) kernel, aka. the power series kernel, see e.g. \citep{Smola00, Zwicknagl09}.   Further, if $n_{b,k}\rightarrow \infty$, then  \eqref{eq:ps_kerenl} converges to 
{\small\begin{align}\label{eq:ps_kerenl_inf}
&\begin{matrix}
{M_{G_k}}^{\text{DP}}(\eta_k)\\
{M_{T_k}}^{\text{DP}}(\eta_k)
\end{matrix}:\left\{
\begin{array}{lll}
&\Gamma_{G_k}^{\text{DP}}= \frac{\alpha_G}{1-\lambda \phi^1  (\phi^1)^T  }+ \frac{\beta_G}{1-\kappa \phi^1 (\phi^1)^T  }\\
& C_{G_k}^{\text{DP}}= \frac{\alpha_G}{1-\lambda \phi^1  (\phi^1)^T  }- \frac{\beta_G}{1-\kappa \phi^1( \phi^1)^T  }\\
\end{array}\right.
\end{align}}\normalsize
where $\eta_k=[\alpha_G, \lambda, \beta_G, \kappa]$, and the convergence holds because all entries of $\lambda^{1/2} \phi^1$ and $\kappa^{1/2} \phi^1$ are within the unit disk. 
\end{Proposition}
Henceforth, the DP kernel is referred to as \eqref{eq:ps_kerenl_inf} by default, and the DP kernel \eqref{eq:ps_kerenl_inf} embeds the prior knowledge of analytic smoothness, i.e., the local function is analytic within the unit disk centering at $j\omega_k$. 

\subsection{Kernel Design for Lightly Damped Systems} \label{subse:kernel_gpr}

On the other hand, the LGPR can immediately use the existing kernels from the global GPR, e.g., the diagonal-correlated (DC) kernel  \citep[Eq. (53)]{LC16}, and the DC kernel plus multiple resonance kernels (DCpRn)  \citep[Eq. (A.3)]{HPJPL22},  with 
$\Gamma_{G_k}(j\omega, j\omega')=C_{G_k}(j\omega, -j\omega')$, 
{\small
\begin{align}
&\Gamma_{G_k}^{\text{DC}}(j\omega, j\omega')=\frac{\lambda}{\sqrt{2\pi}} \frac{1}{\beta\!+\!j\omega\!-\!j\omega'} \bigg(\frac{1}{\alpha\!+\! \beta/2\!+\! j\omega}\!+\! \frac{1}{\alpha\!+\! \beta/2\!-\! j\omega'}\bigg)\label{eq:spe_dc}\\
&\Gamma_{G_k}^{\text{DCpRn}}(j\omega, j\omega')\!=\!\Gamma_{G_k}^{\text{DC}}(j\omega, j\omega')\!+\! \sum_{i=1}^n \Gamma_{G_k}^{\text{R,i}}(j\omega, j\omega'), \label{eq:spe_dcpk}\\
&\Gamma_{G_k}^{\text{R,i}}(j\omega,j\omega')\!=\!\frac{ \gamma_{1,i}^2(j\omega+\beta_{1,i})(-j\omega'+\beta_{1,i})+ \gamma_{2,i}^2\beta_{2,i}^2}{((j\omega+\beta_{1,i})^2+\beta_{2,i}^2)((-j\omega'+\beta_{1,i})^2+\beta_{2,i}^2)},\label{eq:spe_pk1}\\
&\quad \beta_{1,i}>0, \beta_{2,i}, \gamma_{1,i},\gamma_{2,i}\geq 0, i=1, \cdots, n, \nonumber
\end{align}}\normalsize
where $\eta_k =[\lambda,  \alpha, \beta] $ for the DC kernel \eqref{eq:spe_dc} and $\eta_k =[\lambda,  \alpha, \beta, \{\beta_{1,i}, \beta_{2,i}, \gamma_{1,i},\gamma_{2,i}\}_{i=1}^n] $ for the DCpRn kernel \eqref{eq:spe_dcpk}, and \eqref{eq:spe_pk1} is named the resonance kernel with Rn indicating the sum of $n$ resonance kernels.  
It is shown in \cite{HPJPL22} that the  DCpRn kernel \eqref{eq:spe_dcpk}  has a better performance than the DC kernel \eqref{eq:spe_dc}, but it has too many hyper-parameters, leading to difficulties in the hyper-parameter estimation.

However, the local FRF is more likely to have only one resonance, and, by exploiting the partial fractional decomposition (PFD), the vicinity of this resonance can be captured by a second-order resonant system.
Recall that the DP kernel \eqref{eq:ps_kerenl_inf} embeds a stronger smoothness than the DC kernel \eqref{eq:spe_dc}. Then we have the following proposition. 

\begin{Proposition}[DCpR1 Kernel]\label{pr:dcpk1}
The DCpR1 kernel,  \eqref{eq:spe_dcpk} with $n=1$,  can be used in the LGPR to handle lightly damped systems. 
\end{Proposition}
\begin{Proposition}[DPpR1 Kernel]\label{pr:pspk1}
The DPpR1 kernel, i.e., the sum of the DP kernel \eqref{eq:ps_kerenl_inf} and R1 kernel \eqref{eq:spe_pk1}, can be used in the LGPR to handle lightly damped systems. 
\end{Proposition}

 The DC kernel \eqref{eq:spe_dc} and the DP kernel \eqref{eq:ps_kerenl_inf} have close performances for the FRF estimations, while their noise variance estimations may have a large difference. This is related to the overfitting issue:  
 the DC kernel \eqref{eq:spe_dc} can be reduced to an identity kernel $I_{4\ell+2}$ when $\beta\rightarrow 0$ and $\alpha\rightarrow \infty$, implying both the transient term and the noise term in \eqref{eq:map_est_G} are identity kernels $I_{4\ell+2}$.  Consequently, the noise would be mixed with the transient and thus underestimated. In this regard, the DP kernel \eqref{eq:ps_kerenl_inf} is superior to the DC kernel \eqref{eq:spe_dc}.

\section{Simulation}\label{se:sim}
In this section, we run a simple experiment to compare the proposed LPGR to the LPM, LRM, and ILRM. 
\subsection{System and Data Generation}
We consider the testing continuous-time system $G$ from \cite{HPJPL22}, see its true FRF in Fig. \ref{fig:res1}. The input $u(t)$ is a band-limited multisine signal generated by the MUMI toolbox \citep{CPS20}, where the sampling interval is $T_s=0.1$s, the period is 10240 or 3100, and the maximum excited frequency is $0.4*2 \pi/ T_s$ rad/s.  The output $y(t)$ is obtained by the MATLAB function \texttt{y(t)=lsim(G,u(t))+v(t))}, where $v(t)$ is white Gaussian distributed with signal-to-noise ratio (SNR) being 20dB or 60dB. To avoid the initial condition and induce the strong leakage, multiple periods of $u(t)$ and $y(t)$ are run, but only the last $N=10000$ and $N=2500$ samples are collected for period 10240 and 3100, respectively.  The local window size is $2\ell+1$ = 41 for $N$=10000 and $2\ell+1$ = 11 for $N$=2500, which cover the same local FRF with different sampling interval  $2\pi/(NT_s)$.
\subsection{Estimates Setup}
The LPM and LRM use the sample code from \cite{CPS20}, where a truncated SVD is added to calculate the inverse of ill-condition $\Psi_k^H \Psi_k$. 
Assume the same polynomial orders in every local window, i.e., $n_{b,k}=n_{i,k}$ for the LPM and  $n_{a,k}=n_{b,k}=n_{i,k}$ for the LRM. The polynomial orders are fixed to be 2, or selected by the MDL criterion \eqref{eq:mdl}, where these estimates are named LPM(2)/LRM(2), and LPM(MDL)/LRM(MDL), respectively. The ILRM(MDL) minimizes the LOE criterion by MATLAB function \texttt{fmincon()} with the model order and starting point given by the LRM(MDL).

The LGPR estimate is named with its kernel, such as LGPR(DC). The optimization of the hyperparameter tuning \eqref{eq:eb} adopts the MATLAB function \texttt{MultiStart()} with \texttt{fmincon()}, where the number of runs is 5 times the number of hyperparameters.  

The performance of estimates is evaluated by  the mean square error (MSE) in dB, 
{\small\begin{align}
\text{MSE}(\hat G)= \frac{1}{N}\sum\limits_{k\in \Omega^t}(\|\hat G(j\omega_k)- G^o(j\omega_k)\|^2). \label{eq:avg_res}
\end{align}}\normalsize
where $\Omega^t= \{k: 0\leq \omega_k< 2\pi   \}$ is the interest of frequency. 
\subsection{Results and Findings}
\begin{table}[ht]
  \caption{MSE in dB for $\hat G(j\omega)$.}\label{tab:avg_res}
\begin{tabular}{ccccc}
\toprule 
SNR & 20dB & 60dB& 20dB & 60dB\\
\hline
$N(2\ell+1)$ & 2500(11)& 2500(11)& 10000(41)& 10000(41)\\
\midrule
LPM(2)& -40.66& -40.11&-41.85&   -41.93\\
LPM(MDL)& -39.83& -39.69&   -40.17&  -40.10\\
LGPR(DC)&-42.93& -43.42& -91.97&  -79.06\\
LGPR(DP)& -39.53& -39.56& -40.81&  -40.64\\
LRM(2)& \textbf{-93.64}& \textbf{-170.31}& -96.78&  \textbf{-182.08}\\
LRM(MDL)& \textbf{-91.60}& -133.33&-97.46 & \textbf{-182.35}\\
ILRM(MDL)& \textbf{-94.33}& -128.82&\textbf{-111.54} & \textbf{-182.01}\\
 LGPR(DCpR1)& -68.18 & \textbf{-174.62}& \textbf{-113.38}& \textbf{-182.23}\\
 LGPR(DPpR1)& \textbf{-101.49}& \textbf{-170.78}& \textbf{-115.06}& \textbf{-181.95}\\
\bottomrule
  \end{tabular}
 {\small In each column, the values no larger than the lowest values plus 10 are bold since the difference smaller than 10 is negligible.  }
\end{table}
\begin{figure*}
\center
\begin{minipage}[b]{0.3\linewidth}
\includegraphics[width=1\linewidth]{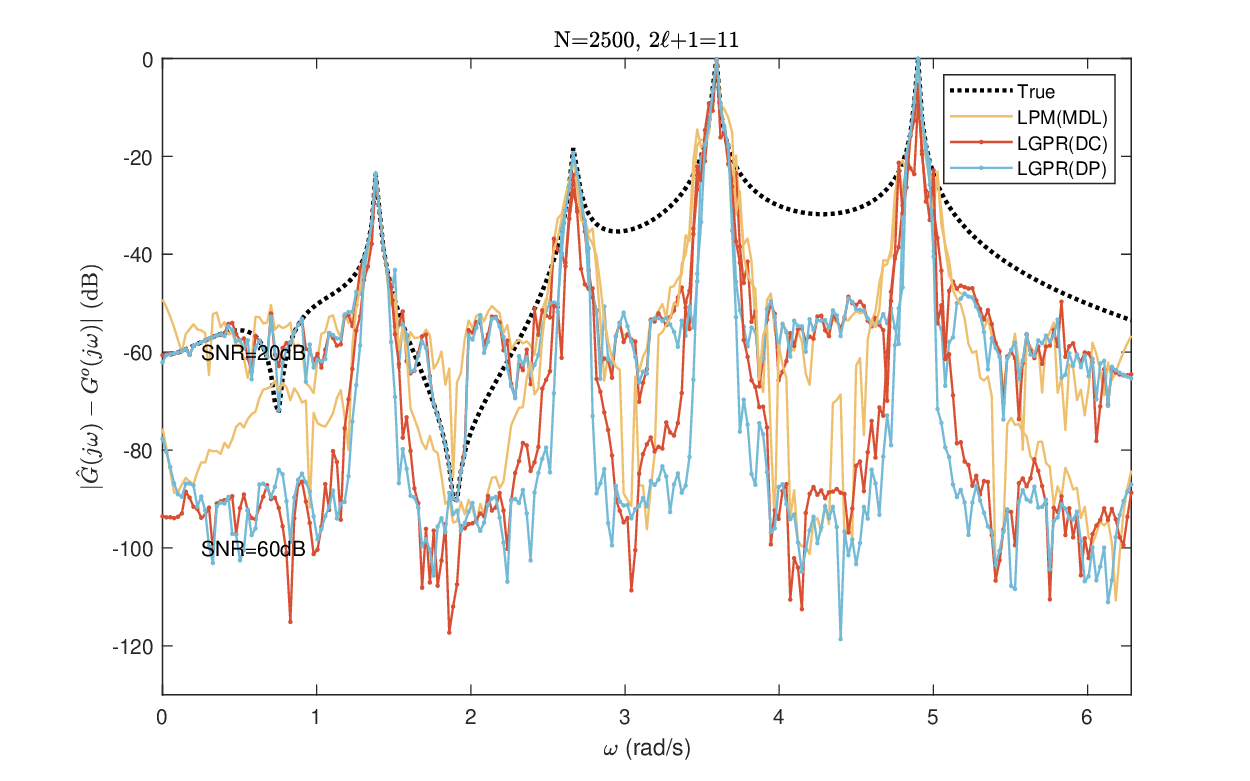}
\end{minipage}
\begin{minipage}[b]{0.3\linewidth}
\includegraphics[width=1\linewidth]{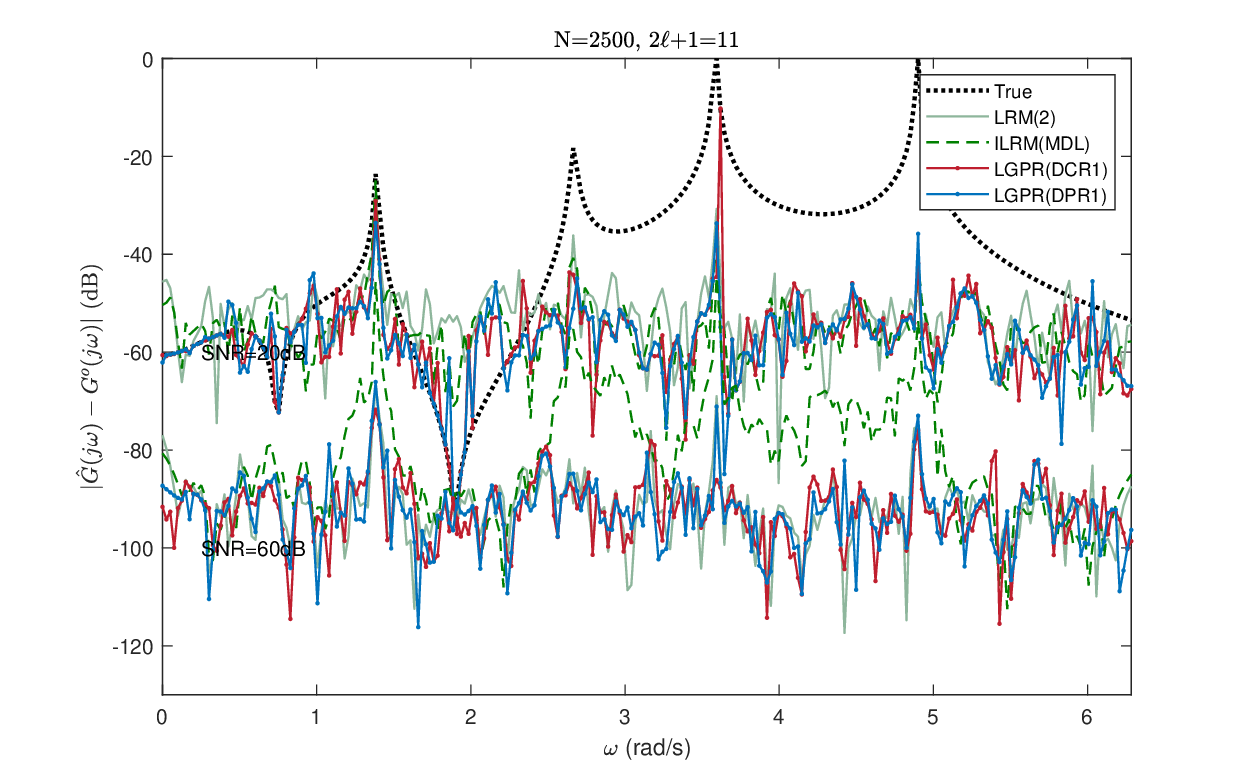}
\end{minipage}
\begin{minipage}[b]{0.3\linewidth}
\includegraphics[width=1\linewidth]{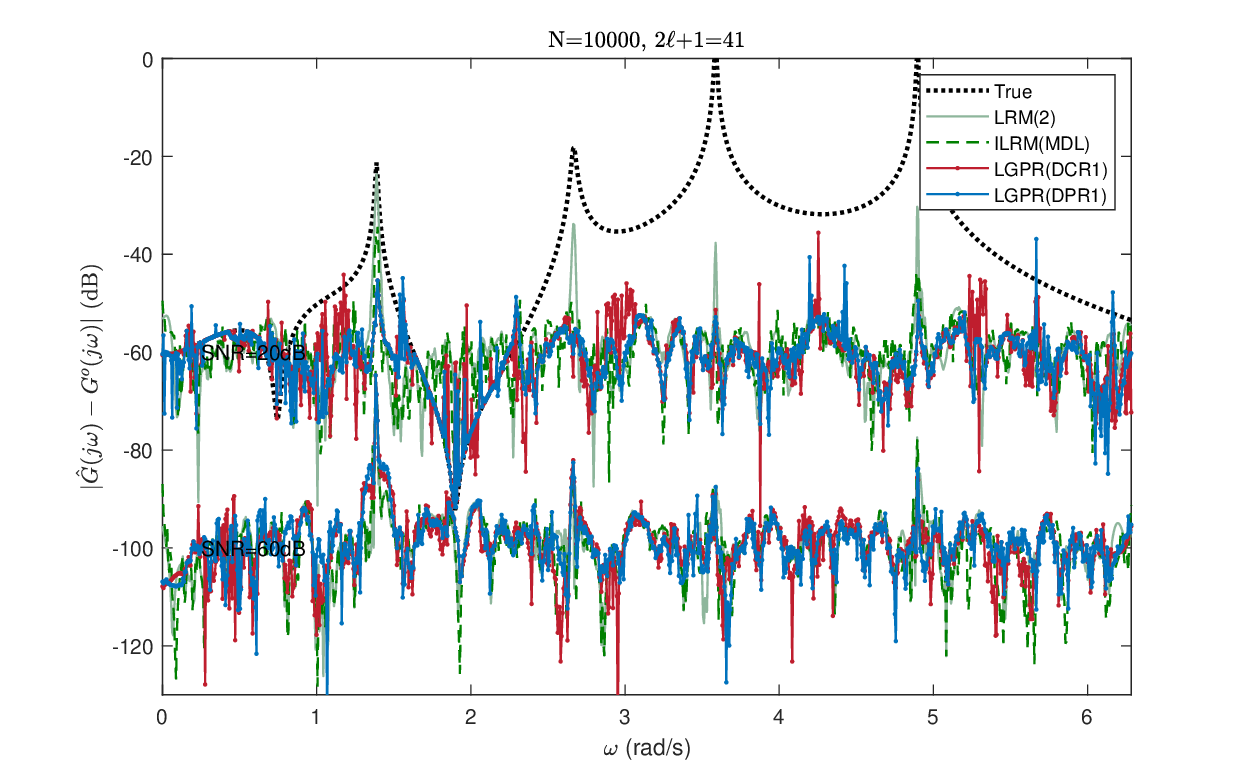}
\end{minipage}
\begin{minipage}[b]{0.3\linewidth}
\includegraphics[width=1\linewidth]{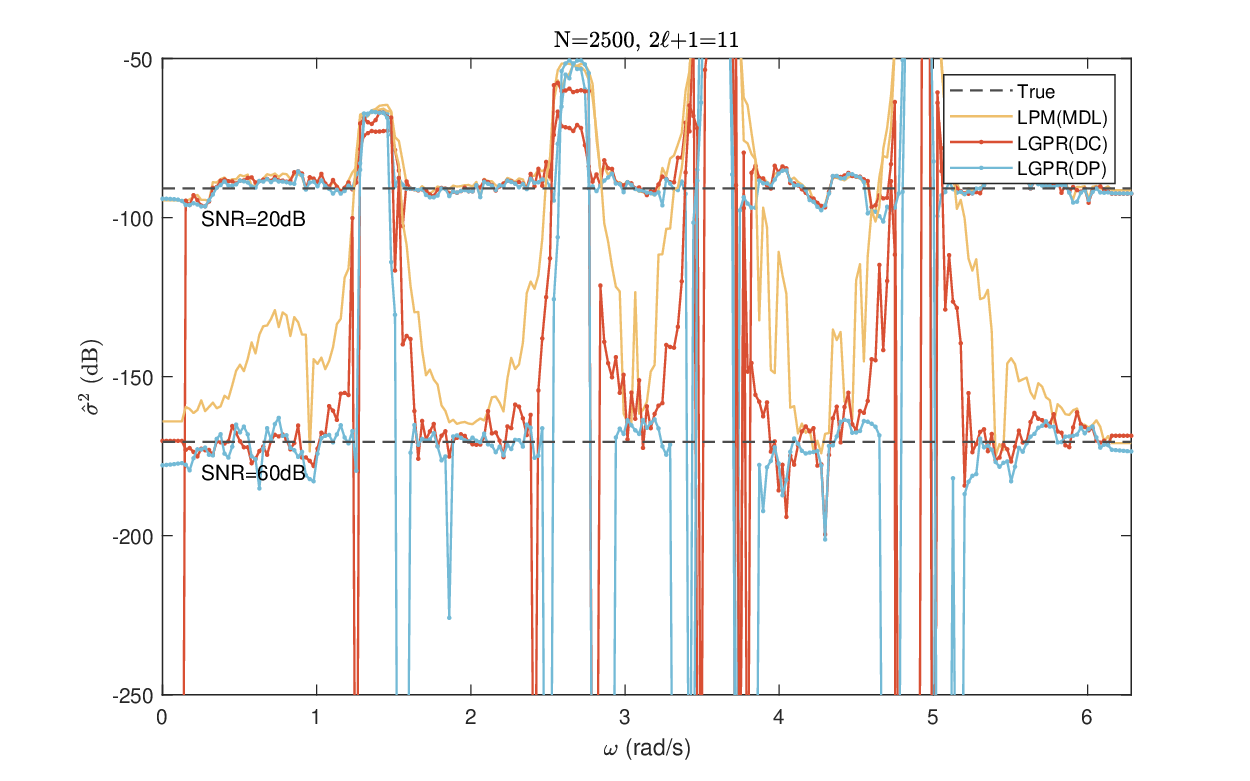}
\end{minipage}
\begin{minipage}[b]{0.3\linewidth}
\includegraphics[width=1\linewidth]{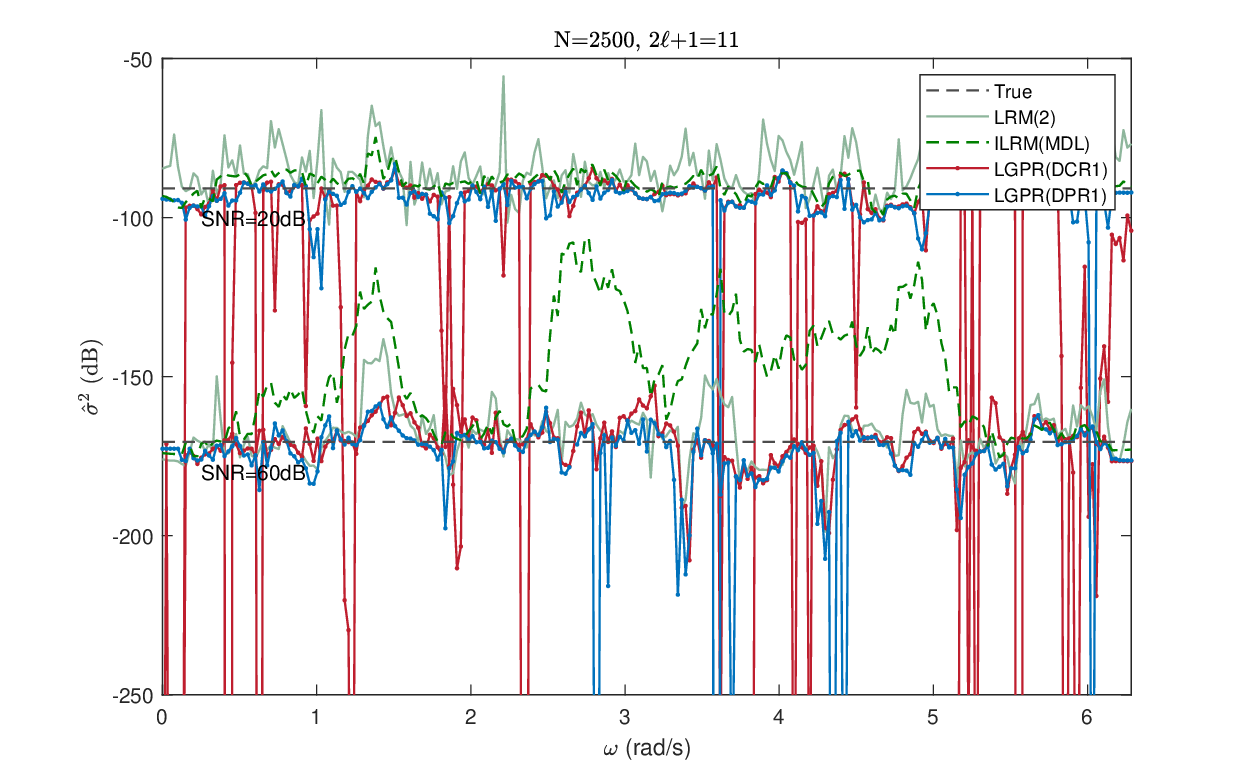}
\end{minipage}
\begin{minipage}[b]{0.3\linewidth}
\includegraphics[width=1\linewidth]{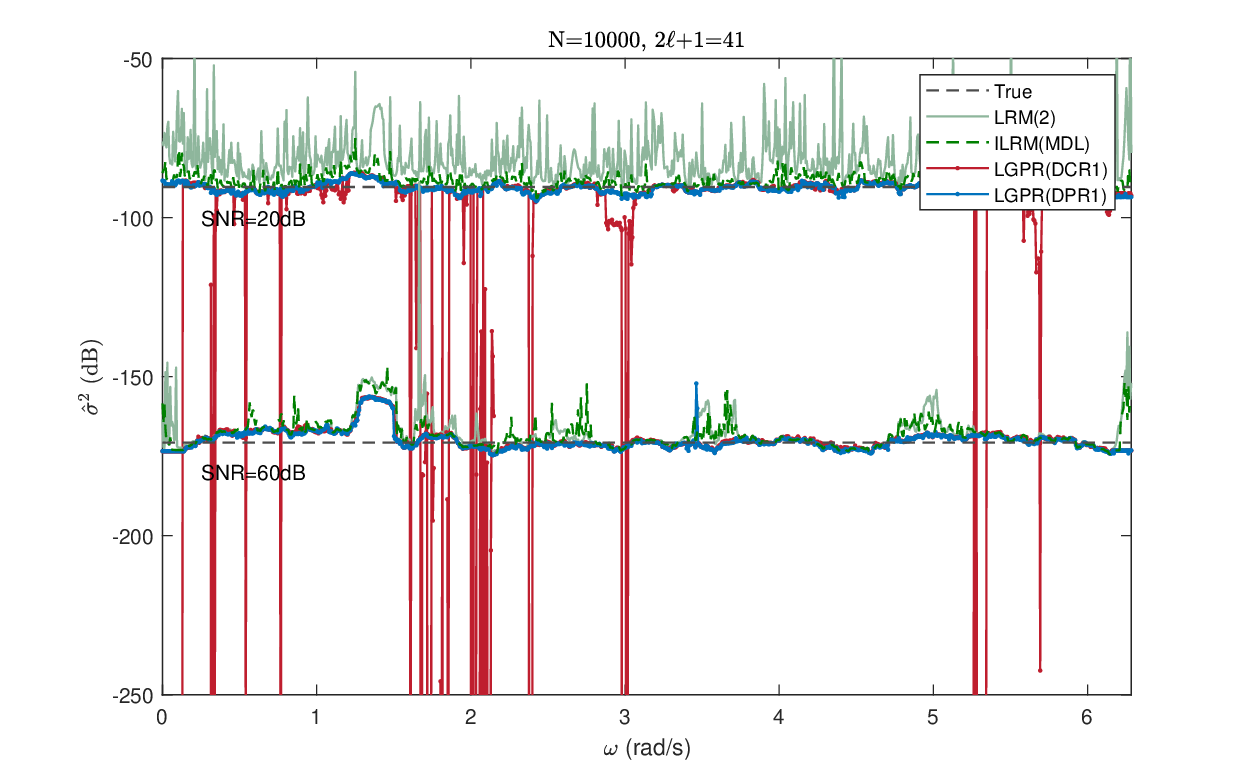}
\end{minipage}
\caption{The residual of the FRF estimates (upper panels) and the noise variance estimates (bottom panels), where the dashed lines are the true FRF and true noise variance.   }  \label{fig:res1}
\end{figure*} 

Table \ref{tab:avg_res} shows the MSE of the FRF estimates, and
Fig. \ref{fig:res1} shows the residual of some selected FRF estimates, $|\hat G(j\omega)- G^o(j\omega)|$,  and their noise variance estimates, $\hat{\sigma}^2$.  The simulations results show that:
\begin{itemize}
\item The LPM,  LGPR(DC), and  LGPR(DP) have poor FRF estimation accuracy around the resonance due to the false of Assumption \ref{as:analytic}. 
\item The LRM(2), LRM(MDL), and ILRM(MDL) do not achieve the best FRF estimation MSE (bold values) for all columns in Table \ref{tab:avg_res}. 
Specifically, 
\begin{itemize}
    \item[-] The worse performances of LRM(MDL) and ILRM(MDL) in N=2500 and SNR=60dB are due to the overfitting when the sample size is small. 
        \item[-] The worse performances of LRM(2) and ILRM(2) in N=10000 and SNR=20dB are due to the biased noise variance estimations, see Fig. \ref{fig:res1}. 
\end{itemize}
\item The  LGPR (DCR1) has the overfitting issue in the noise variance estimate, which is illustrated as  spikes in Fig. \ref{fig:res1}. It is shown that this overfitting may affect the FRF estimation MSE.  
\item The  LGPR (DPR1)  achieves the best FRF estimation MSE (bold values) for all columns in Table \ref{tab:avg_res}. Its noise variance estimate has spikes for a small sample size but does not for a large sample size.  The spikes do not affect the FRF estimation MSE. 
\end{itemize}
\section{Conclusion}
This work proposes a local Gaussian process regression and new kernels for FRF estimation.  The proposed local Gaussian process regression (LGPR) provides a new route to tackle the model order selection and the identification of lightly damped systems. It is shown in the simulations that the LGPR has the best FRF estimation accuracy in terms of the mean square error (MSE), and the LGPR is more robust to different sample sizes and noise levels.
\bibliography{database}             
                                                   







\appendix

\end{document}